\documentclass[conference]{IEEEtran}
\IEEEoverridecommandlockouts
\usepackage{cite}
\usepackage{amsmath,amssymb,amsfonts}
\usepackage{algorithm}
\usepackage{algpseudocode}
\usepackage{graphicx}
\usepackage{textcomp}
\usepackage{listings}
\usepackage{xcolor}
\usepackage{listings}
\usepackage{float}
\usepackage{caption}
\def\BibTeX{{\rm B\kern-.05em{\sc i\kern-.025em b}\kern-.08em
    T\kern-.1667em\lower.7ex\hbox{E}\kern-.125emX}}
\begin{document}

\title{Predictive Power of LLMs in Financial Markets}

\author{\IEEEauthorblockN{Jerick Shi}
\and
\and
\IEEEauthorblockN{Burton Hollifield}
}

\maketitle

\begin{abstract}
Predicting the movement of the stock market and other assets has been valuable over the past few decades. Knowing how the value of a certain sector market may move in the future provides much information for investors, as they use that information to develop strategies to maximize profit or minimize risk. However, market data are quite noisy, and it is challenging to choose the right data or the right model to create such predictions. With the rise of large language models, there are ways to analyze certain data much more efficiently than before.

Our goal is to determine whether the GPT model provides more useful information compared to other traditional transformer models, such as the BERT model. We shall use data from the Federal Reserve Beige Book, which provides summaries of economic conditions in different districts in the US. Using such data, we then employ the LLM's to make predictions on the correlations. Using these correlations, we then compare the results with well-known strategies and determine whether knowing the economic conditions improves investment decisions. We conclude that the Beige Book does contain information regarding correlations amongst different assets, yet the GPT model has too much look-ahead bias and that traditional models still triumph.
\end{abstract}


\section{Introduction}
Predicting whether the stock market would go up or down has always been a challenge for investors. Researchers have tried several different methods in order to predict market movement, ranging from different statistical machine learning models to trends on social media, the goal being to find the optimal strategy to make the most amount of money. With the strong capabilities of large language models, it begs the question of whether we can improve investing using such tools. In other words, our goal is to determine whether we can predict people's beliefs on the market through the use of LLM's on different economic sources. 

\subsection{Challenges}
\begin{figure}[htbp]
\centerline{\includegraphics[scale=0.3]{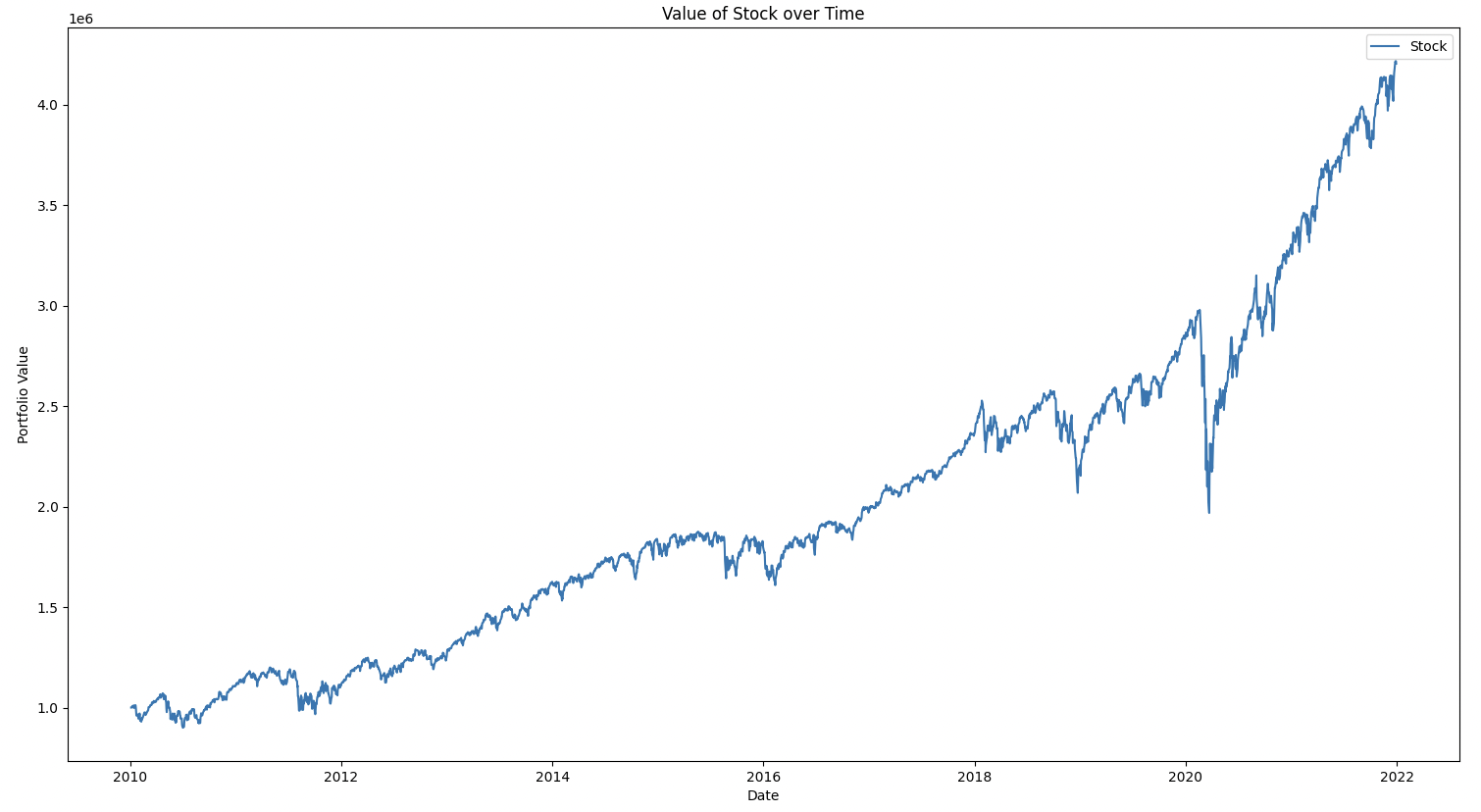}}
\caption{Movement of Stocks over Time}
\label{stock}
\end{figure}

There are several challenges that occur when trying to predict the movement in the market. For one, market data are extremely noisy, as can be seen in Figure \ref{stock}. The market is extremely volatile and reacts quickly to news. Furthermore, any significant events that occur in the world can have a huge impact on the market, such as elections, technological innovations, or pandemics.  Due to the large number of market makers, prices can also change drastically each day. Hence, making predictions in the market is quite a daunting task.\\

Bescause market data are noisy, it is highly likely that any model will overfit the old data. Models are highly likely to capture insignificant information, making it difficult to generalize for future times. For example, if you were to predict how the market moves in the early twentyfirst  century, it is highly unlikely that the model will be able to predict that Covid would happen, in which stocks plummeted. Hence, even though there are several models that performed well in previous years, few models perform as well in future times.\\

Not only is quantitative data extremely volatile, market news is quite complex as well. Although news data is quite abundant, it is challenging to choose which sources of data reflect the current market. News sources often have their own opinions on how the market moves, so focusing on a single news source might not best reflect the entire market.\\

Hence, due to the amount of noise contained in both quantitative and qualitative data, predicting market movement has been a huge challenge. Our approach addresses these challenges by using large language models on much cleaner datasets.
\subsection{Challenges with LLM's}
Large language models, while powerful, also pose several challenges. For one, the output itself is a probability distribution. Depending on how likely the final outputs are, since most LLM's are closed-source, the experiments aren't perfectly replicable. Furthermore, LLM's are highly likely to hallucinate, returning either answers that are not correct or even providing responses that do not fit into the context of the problem. Finally, because large language models are black boxes, it is quite challenging to determine which parts of the input are useful for its output.
\subsection{Related Works}
Several research papers have used similar ideas of using sentimental analysis in news articles to predict market movement. In particular, Kalyani (2016) \cite{b1} introduced the idea of using news sentiment analysis to predict stock trends using machine learning techniques. However, these techniques are not always successful in extracting non-linear information. Ren (2022) \cite{b2} introduces more advanced techniques to analyze the impact of news on the change in stock price. In particular, a bidirectional LSTM model was created. However, the embedding model is a word2vec model which was created in 2013. Although functional, such models may not always capture the significant information contained within longer articles, which newer embedding models do a much better job of.\\

Perhaps one of the most recent papers relating to the use of LLM's on predicting market data is by Bybee (2023) \cite{b3}, where he measures economic sentiment by applying LLM's on news articles. We intend to improve on the basis of his approach through several different methods. For one, Bybee attempts to predict expected returns, which is quite challenging because returns are so noisy. Rather, we will predict how different stocks and bonds correlate with one another, as well as predict how their prices vary over time, which we use to construct our optimal portfolio. Furthermore, Bybee uses news article data, which changes frequently and is noisy. We intend to utilize the Beige Book, which contains the summaries of the economic conditions of different states in the United States. More importantly, it is published every 1-2 months, so it provides much cleaner data to condition on.

\subsection{Contributions}
From this report, we make 3 main contributions. First, we show that regardless of how well the prompt is, there is always an unavoidable look-ahead bias when using the GPT model. Secondly, we show that certain inputs to the prompt such as specific numerical data do not necessarily improve prediction results. Finally, we show that, with regard to the optimal portfolio, using information from the BERT model still does much better than using information from the GPT model.

\section{Methodology}
\subsection{Data}
We obtain historical stock prices directly from Yahoo Finance, where we take the prices from $\text{\^{}} GSPC$ from the first day of 1985 to the last day of 2023. The bond data comes from the same source under the ticker $AGG$, but the data only span from the middle of 2003 to the end of 2023.\\

Since there are no open sources for earlier prices, then go to Wharton Research Data Services and the FRED, where we were able to get 1-year, 2-year, 5-year, 7-year, 10-year, 20-year, 30-year bill returns as well as the ICE BofA US Corporate Index Total Return Index Value. These data go back to 1980. We then run a linear regression algorithm as shown in Algorithm 1.

\begin{algorithm}
\caption{Bond Regression Algorithm}\label{alg:cap}
\begin{algorithmic}
\State $bond\_data$ $\gets$ $AGG$ from 2003 to 2023
\State $variables$ $\gets$ [1-year, 2-year, 5-year, 7-year, 10-year, 20-year, 30-year, corporate] bill returns
\For{each $combination$ of $variables$}
\State Fit linear regression of $combination$ on $bond\_data$ to predict AGG (split 80/20)
\State Calculate out-of-sample $R^2$
\EndFor
\State return best combination + best $R^2$
\end{algorithmic}
\end{algorithm}

\begin{figure}[h]
    \centering
    \includegraphics[width=0.45\textwidth]{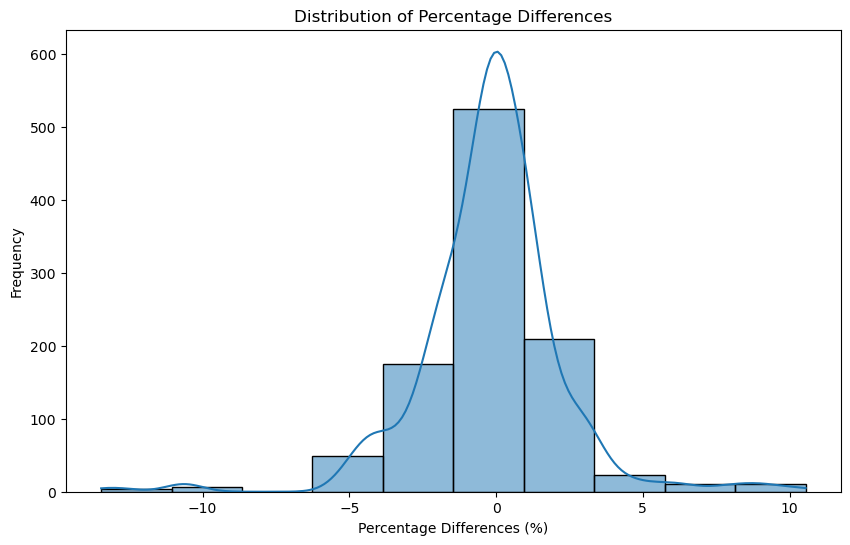}
    \caption{Error Distributions of Predicted Prices}
    \label{error}
\end{figure}

Our best result chooses the corporate index, 1-year, 2-year, 5-year, 7-year, and 10-year bill returns with an out-of-sample $R^2$ of $\bf{0.9901}$. We show the errors out-of-sample in Figure \ref{error}.

\begin{figure}[h]
    \centering
    \includegraphics[width=0.45\textwidth]{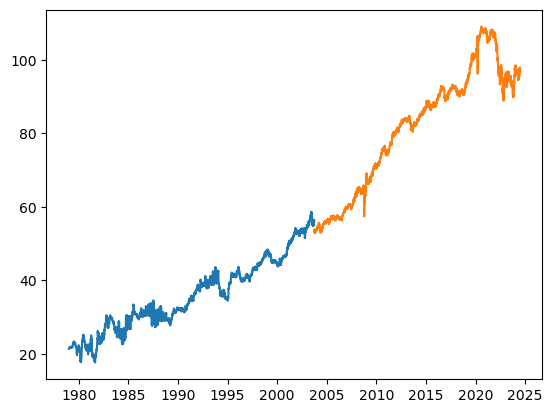}
    \caption{Predicted Prices (Blue) of Bonds}
    \label{ppb}
\end{figure}

We observe that since the errors are all quite small, so our model should produce relatively accurate prices for AGG. We also show in figure \ref{ppb} predictions for earlier AGG prices (blue being our prediction, orange being the actual prices).\\

Finally, the beige book is publicly available on the Federal Reserve Bank of Minneapolis. Each year, 5 months per year, and 12 states for each month, an article is published. Since the months and states are always the same per year, the articles are quite easy to scrap.

\subsection{Correlations}
The actual correlations are defined by taking the daily percent changes for stocks and bonds in a month and then calculating their correlation. We define that to be the 'true' correlation of stocks and bonds for that month.\\

For the Beige Book, we shall use the GPT model in order to obtain the correlations. The model we shall use is the GPT-3.5 Turbo Model, which is trained with data up until September 2021. We also set the temperature to 0 so that it is most likely to be replicable. We ask it using the prompt in figure \ref{prompt}.\\

\begin{figure}[h]
    \centering
    \begin{lstlisting}
You are a financial agent in the
year {year}. You don't know anything
that happens during this year 
or anything afterwards. The following
data is an article regarding the 
economic conditions: {article}

Furthermore, previous {num_correlations}
month's correlations of stocks and bonds
are given by {correlations} from
oldest to most recent.

Only using the above data and nothing
else, how do you think stock returns
and bond returns will be correlated?

Do not use any other outside
information. You must choose a side,
you don't need to be certain about it.

Respond with only 0, 1 or 2,
0 being negatively correlated, 1 being
uncorrelated, and 2 being positively
correlated.

You should return 1 digit and
nothing else.

Example: 1. This example implies
the guess is positive correlation.
Ensure the formatting is correct.
    \end{lstlisting}
    \caption{Example Prompt for GPT}
    \label{prompt}
\end{figure}

Another way we shall ask for it is in bins, where, rather than returning numbers between 0 and 2, we can ask for numbers between 0 and 10 to be more specific.\\

With each of those answers, the model also outputs a probability $p$, with respect to the likelihood. Since temperature=0, it will always choose the most likely one. We can then multiply that probability by the correlation, so that the correlations are a bit more stable. In order to make the probabilities less extreme, we also try to multiply the correlation by $p-(1-p)$, which we call the scaled strength. For the Beige Book, each month has several articles, so we shall take the average amongst all of the articles.\\

There are also certain cases where the article may exceed the context length. In that case, we either segment it into different parts to feed into the model, and then take the average, or if there are more than 10 parts, we randomly take 10 subsections, and then take the average of those.\\

Hence, we have two different ways to ask for correlations, and each way then has three different correlations that we can use.

\subsection{Bert Correlations}
In addition to the GPT models, we also wanted to determine whether the BERT model would also function in analyzing data sources. In this case, we train the BERT model to output values similar to those of the GPT models. For the first case, we take each of the actual correlations, round it to $-1$, $0$, or $1$, and then treat it as a classification algorithm. For the case with bins, we also split it into $11$ different classes depending on its correlations, and those outputs are then the labels for each of the articles in the Beige Book for a month.\\

The model is then trained on all data from 1980 to 2021 September, similarly to the GPT model, where we train for 10 epochs. The predicted values are then calculated similarly to the GPT model, where each value also had a strength value. Hence, the BERT model also created 3 different versions for each way of calculating correlations.

\subsection{All Correlations}
To summarize, here are all the correlations that we use in our experiments:
\begin{itemize}
  \item Beige Correlations Original (3 versions)
  \item Beige Correlations with Bins (3 versions)
  \item BERT Correlations Original (3 versions)
  \item BERT Correlations with Bins (3 versions)
\end{itemize}

\section{Experiments}
In this section, our aim is to answer 4 different questions:
\begin{itemize}
  \item Is there evidence of look-ahead bias in the GPT-3.5 model?
  \item Does adding historical correlations improve predictions?
  \item Does the GPT-3.5 model perform better than the BERT model in analyzing federal data?
  \item Is the GPT model a more effective way to make money than standard models?
\end{itemize}

\subsection{Hypothesis Testing}
For the first three questions, we can use hypothesis testing, where we can determine whether the errors increase or decrease significantly. We shall analyze the RMSE's of the predicted and actual monthly correlations, which is defined as
$$RMSE=\sqrt{\sum_{i=1}^n(y_i-\hat{y_i})^2}$$
where we let the number of months $n$ be a hyperparameter. In order to maintain independence, each $RMSE$ will span $n$ months after the end of the previous $n$ months. Furthermore, since the GPT-3.5 model is 'trained up to September 2021', we shall use 1980-01 to 2021-09 as the training set and 2021-10 to 2024-06 as the testing set. 
\subsection{Simulation}
In order to determine whether GPT models are beneficial for use in the real world, we shall simulate how portfolio values change over time through different strategies using the GPT model.\\

We use 2 main metrics to measure performance: The PnL and sharpe ratio. PnL refers to how much the value of the portfolio changes following a specific strategy. Hence, PnL provides a useful measurement since it is the easiest representation of how much money you would gain or loss over time. The Sharpe ratio for asset or strategy $a$, $\theta_a$, is 
\begin{equation}
\theta_a=\frac{E[R_a-R_f]}{\sigma_a}\label{eq1},
\end{equation}
where $R_a$, $R_f$  are the returns of asset $a$ and the risk-free rate respectively, and $\sigma_a$ is the standard deviation of the returns.

We  first test it on the 2-variable minimization strategy, where given the correlation and variances of stocks and bonds, there exists an allocation of weights that minimizes risk. The good thing about this strategy is how simple it is. Since there are only 2 variables, there exists a closed-form solution for the weights given the different parameters.\\

Specifically, our goal is to choose a $w$ such that putting $w$ of our portfolio in stocks and $1-w$ of our portfolio in bonds minimizes variance. Mathematically, 
\begin{equation}
w^*={argmin}_w\:  Var(wS+(1-w)B)\label{eq1}
\end{equation}
We then can then minimize the variance as follows:
\begin{align*}
&\frac{\partial}{\partial w}Var(wS+(1-w)B)=0\\
&\implies \frac{\partial}{\partial w}[w^2\sigma_S^2+2w(1-w)\rho\sigma_S\sigma_B+(1-w)^2\sigma_B^2]=0\\
&\implies 2w\sigma_S^2+2(1-2w)\rho\sigma_S\sigma_B)-2(1-w)\sigma_B^2=0\\
&\implies w(2\sigma_S^2-4\rho\sigma_S\sigma_B+2\sigma_B^2)=2\sigma_B^2-2\rho\sigma_S\sigma_B\\
\end{align*}
\begin{equation}
\implies w^*=\frac{\sigma_B^2-\rho \sigma_S\sigma_B}{\sigma_S^2+\sigma_B^2-2\rho\sigma_S\sigma_B}\label{eq2}
\end{equation}
where $\sigma_S$ and $\sigma_B$ represent the standard deviations of stocks and bonds respectively, and $\rho$ corresponds to their correlation. Hence, we can keep the standard deviations the same for all strategies by using the exponential moving averages and then seeing how the different correlations might affect the weights.\\

We then consider the case of using several more asset classes. In addition to stocks and bonds, we shall use commodities ($\text{\^{}}SPGSCI$), real estate ($\text{\^{}}SP500-60$), and dollar value ($DX-Y.NYB$). Although the training set may decrease due to some of the indices that exist later, they still provide a good representation of the performance of portfolios before and after Covid. Our calculations will also be a bit different, since we need to introduce some linear algebra. Let $\bf{w}$ be the vector of portfolio weights and $\bf{\Sigma}$ be the covariance matrix. The problem then becomes:
\begin{align*}
    \text{min}_{\bf{w}}&\: \bf{w}^T\bf{\Sigma}\bf{w}\\
    \text{s.t.}&\:\bf{1}^T\bf{w}=1
\end{align*}
Introducing the Lagrangian, we have that
$$\mathcal{L}(\bf{w},\lambda)=\bf{w}^T\bf{\Sigma}\bf{w}-\lambda(\bf{1}^T\bf{w}-1)$$
We then calculate the optimal weight allocation below:
\begin{align*}
    \frac{\partial \mathcal{L}}{\partial \mathbf{w}}&=2\bf{\Sigma}\bf{w}-\lambda\bf{1}=0\implies 2\bf{\Sigma}\bf{w}=\lambda\bf{1}\\
    \frac{\partial \mathcal{L}}{\partial \lambda}&=-(\bf{1}^T\bf{w}-1)=0\implies \bf{1}^T\bf{w}=1\\
   & \implies \bf{w}=\frac{\lambda}{2}\bf{\Sigma}^{-1}\bf{1}\\
    & \implies \frac{\lambda}{2}\bf{1}^T\bf{\Sigma}^{-1}\bf{1}=1\\
     &\implies\lambda=\frac{2}{\bf{1}^T\bf{\Sigma}^{-1}\bf{1}}
\end{align*}
\begin{equation}
\implies w^*=\frac{\Sigma^{-1}\bf{1}}{\bf{1}^T\bf{\Sigma}^{-1}\bf{1}}
\end{equation}
Using all of those parameters, we then demonstrate our strategy in Algorithm 2. Furthermore, we shall split the periods into 2, pre-Covid and post-Covid, to determine how the models behave over different scenarios. 

\begin{algorithm}
\caption{Simulation Algorithm for Porfolio Value}\label{alg:cap}
\begin{algorithmic}
\State $finance\:data\gets yfinance(start\:date, end\:date)$
\State $beige\:data\gets federalreserve(start\:date, end\:date)$

\For{day/month  in  data}
    \State $porfolio[month]\gets Update(portfolio[month-1])$
    \If{firstTradingDay(day,month)}
    \State $\sigma_{S}\gets rollingAverage(month, Stock)$
    \State $\sigma_B\gets rollingAverage(month, Bond)$
    \State $\rho_{month} \gets askGPT(beige\:data[month],\rho_{-3::-1})$
    \State $w_S \gets minimizeVariance(\sigma_B,\sigma_S,\rho_{month})$
    \State $w_B\gets 1-w_S$
    \State $w_S',w_B'\gets scaled(w_S,w_B,\sigma_B)$ 
    \State $w_I\gets 1-w_S'-w_B'$
    \EndIf
\EndFor

return $portfolio$
\end{algorithmic}
\end{algorithm}

\section{Results of Hypothesis Tests}
\label{res}
\subsection{Look-Ahead Bias}

If a model has look-ahead bias, that would imply that the model has prior knowledge regarding the future within the training process. Hence, if it was given information from May, and is asked to predict information in June, since the model is trained on information including that in June, there is a possibility that the model doesn't solely use the information in May in order to predict.\\

Hence, let $e_{train}$ and $e_{test}$ be the RMSE of the training and testing set respectively for the GPT model. We shall test the following hypothesis:
$$H_0:\, e_{train}=e_{test}$$
$$H_a:\, e_{train}<e_{test}$$
In other words, the null hypothesis is that there is no look-ahead bias, so there should be no different in error. If the error has a significant increase out of sample, that indicates that there is evidence of look-ahead bias.\\

We shall perform a one-sided t-test between 2 sets, the training set and the test set. We shall take the RMSE every $n$ months, where each error is the difference between the true and actual correlation per month, where as $n$ increases, the values become less reliable, since we only have a little less than 3 years in the testing set. We show the results in tables \ref{t1} and \ref{t2}:

\begin{table}[h]
\centering
\begin{tabular}{||c c c c||} 
 \hline
 $n$ & p-value (v1) & p-value (v2) & p-value (v3)\\ [0.5ex] 
 \hline\hline
 1 & 0.121 & 0.083 & \bf{0.044} \\ 
 \hline
 3 & 0.087 & 0.080 & 0.059  \\
 \hline
 6 & 0.173 & 0.158 & 0.140  \\
 \hline
 12 & 0.122 & 0.125 & 0.138\\
 \hline
\end{tabular}
\caption{P-values for Look Ahead Bias (Original)}
\label{t1}
\end{table}

\begin{table}[h]
\centering
\begin{tabular}{||c c c c||} 
 \hline
 $n$ & p-value (v1) & p-value (v2) & p-value (v3)\\ [0.5ex] 
 \hline\hline
 1 & \bf{0.001} & \bf{0.011} & \bf{0.025}\\ 
 \hline
 3 & \bf{0.003} & \bf{0.010} & \bf{0.023}  \\
 \hline
 6 & \bf{0.011} & \bf{0.026} & 0.063  \\
 \hline
 12 & 0.115 & 0.121 & 0.110\\
 \hline
\end{tabular}
\caption{P-values for Look Ahead Bias (with bins)}
\label{t2}
\end{table}

From the above results, we see that while the values for the original strategy do not seem as significant, the values for the strategy with bins seem quite significant, indicating that there is evidence that the errors increase in the testing set. Hence, it is quite likely that when asked questions that are more specific, the model isn't using as much information from the beige book, but rather looking at historical data to make its predictions. For the original beige book strategy, while the results aren't as significant because the model does have access to past returns. 

\subsection{Adding Historical Correlations}
We shall then test if the errors decrease by adding the previous 3 month correlations.\\

Let $e_{wihout}$ and $e_{with}$ be the RMSE of the prompts without or with the past correlations respectively for the GPT model. We shall test the following hypothesis:
$$H_0:\, e_{without}=e_{with}$$
$$H_a:\, e_{without}>e_{with}$$

The testing will be almost the same as before, also using the t-test, except that significant results occur if the errors decrease. We show the results in Tables \ref{t3} and \ref{t4}:

\begin{table}[h]
\centering
\begin{tabular}{||c c c c||} 
 \hline
 $n$ & p-value (v1) & p-value (v2) & p-value (v3)\\ [0.5ex] 
 \hline\hline
 1 & 1.000 & 1.000 & 1.000\\ 
 \hline
 3 & 1.000 & 1.000 & 1.000  \\
 \hline
 6 & 1.000 & 1.000 & 1.000 \\
 \hline
 12 & 0.998 & 1.000 & 1.000\\
 \hline
\end{tabular}
\caption{P-values for historical correlations effect (Original)}
\label{t3}
\end{table}

\begin{table}[h]
\centering
\begin{tabular}{||c c c c||} 
 \hline
 $n$ & p-value (v1) & p-value (v2) & p-value (v3)\\ [0.5ex] 
 \hline\hline
 1 & 0.368 & 0.411 & 0.461\\ 
 \hline
 3 & 0.371 & 0.404 & 0.444  \\
 \hline
 6 & 0.416 & 0.411 & 0.453 \\
 \hline
 12 & 0.443 & 0.431 & 0.467\\
 \hline
\end{tabular}
\caption{P-values for historical correlations effect (with bins)}
\label{t4}
\end{table}

Where we notice that none of the values are close to being significant. In fact, for the original correlations, there is evidence that the guesses become worse by adding noisy data. Hence, we conclude that the GPT model cannot perfectly capture patterns from noisy numerical data.

\subsection{Whether GPT model beats BERT}
Since both the GPT model and BERT model are trained on the same training set, we then want to see whether the GPT model might be better at analyzing federal data than the BERT model. We should split the experiments to the training and testing set, where the first table contain the original correlations, and the second table contain the correlations with bins.

\begin{table}[h]
\centering
\begin{tabular}{||c c c c||} 
 \hline
 $n$ & p-value (v1) & p-value (v2) & p-value (v3)\\ [0.5ex] 
 \hline\hline
 1 & $\mathbf{3.38\times 10^{-6}}$ & \bf{0.041} & 0.259\\ 
 \hline
 3 & $\mathbf{1.12\times 10^{-5}}$  & \bf{0.027} & 0.256  \\
 \hline
 6 & $\mathbf{1.06\times 10^{-4}}$ & \bf{0.031} & 0.253\\
 \hline
 12 & \bf{0.002} & 0.062 & 0.280\\
 \hline
\end{tabular}
\caption{P-values for GPT vs BERT (Training Original)}
\label{t5}
\end{table}

\begin{table}[h]
\centering
\begin{tabular}{||c c c c||} 
 \hline
 $n$ & p-value (v1) & p-value (v2) & p-value (v3)\\ [0.5ex] 
 \hline\hline
 1 & 1.000 & 1.000 & 1.000\\ 
 \hline
 3 & 1.000 & 1.000 & 1.000  \\
 \hline
 6 & 1.000 & 1.000 & 1.000 \\
 \hline
 12 & 1.000 & 1.000 & 1.000\\
 \hline
\end{tabular}
\caption{P-values for GPT vs BERT (Training with Bins)}
\label{t6}
\end{table}

\begin{table}[h]
\centering
\begin{tabular}{||c c c c||} 
 \hline
 $n$ & p-value (v1) & p-value (v2) & p-value (v3)\\ [0.5ex] 
 \hline\hline
 1 & \bf{0.019} & 0.059 & 0.194\\ 
 \hline
 3 & \bf{0.023} & \bf{0.038} & 0.082  \\
 \hline
 6 & 0.085 & 0.132 & 0.219\\
 \hline
 12 & 0.063 & 0.068 & 0.062\\
 \hline
\end{tabular}
\caption{P-values for GPT vs BERT (Test Original)}
\label{t7}
\end{table}

\begin{table}[h]
\centering
\begin{tabular}{||c c c c||} 
 \hline
 $n$ & p-value (v1) & p-value (v2) & p-value (v3)\\ [0.5ex] 
 \hline\hline
 1 & 0.991 & 0.940 & 0.624\\ 
 \hline
 3 & 0.958 & 0.936 & 0.620  \\
 \hline
 6 & 0.923 & 0.938 & 0.607\\
 \hline
 12 & 0.904 & 0.957 & 0.726\\
 \hline
\end{tabular}
\caption{P-values for GPT vs BERT (Test with bins)}
\label{t8}
\end{table}

In tables \ref{t5} and \ref{t7}, we observe that the GPT model seems to have some evidence of performing better than the BERT model in-sample, especially for the original correlation. However, out-of-sample, as shown in tables \ref{t6} and \ref{t8}, it seems that the BERT model performs much better. This may be due to the GPT model having look-ahead bias and being overfit on training data, so it doesn't generalize as well in the testing set. Hence, this demonstrates that while the GPT model is quite a powerful tool in obtaining past specific information, it doesn't generalize well for information it hasn't seen.\\

We also show a visualization of the errors in section \ref{FirstAppendix} in the appendix.

\section{Results of Simulations}
\subsection{2 Variables}
For the simulation, we shall compare amongst these 3 models:
\begin{itemize}
    \item Baseline: Exponential Average Covariances
    \item BERT: Original Correlations V3
    \item GPT: Original Correlations V3
\end{itemize}
For the baseline model, we take in the exponential rolling covariances of stocks and bonds as the time series, since it is able to capture a bit more information than simple rolling average.

The reason we choose these BERT and GPT models is that V3 seems to produce a more accurate correlation than both $V1$ and $V2$, and then the original correlations seem to have less look-ahead bias. Furthermore, we want to determine more on whether GPT or BERT performs better by keeping most of the parameters constant. We show how the portfolio values change over time in figures \ref{s1} and \ref{s2}, and show the final sharpe ratios in table \ref{t9}.

\begin{figure}[h]
    \centering
    \includegraphics[width=0.45\textwidth]{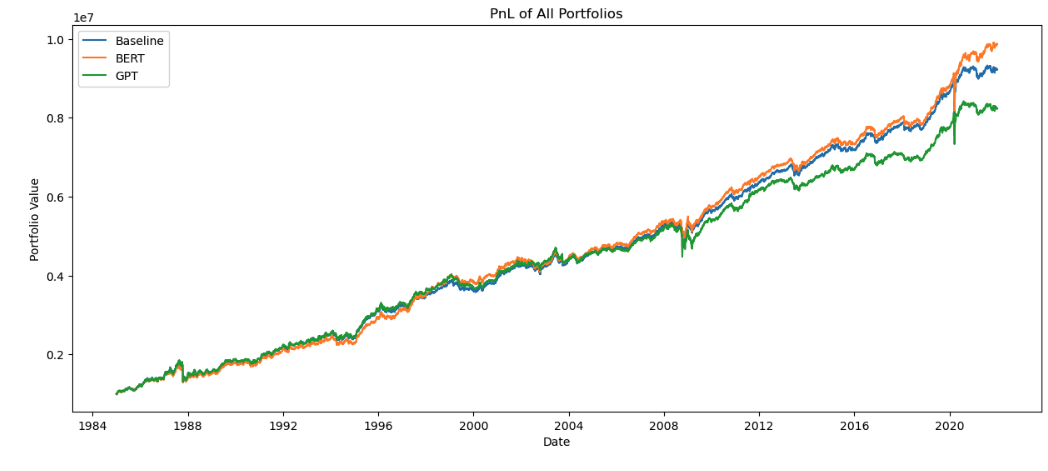}
    \caption{2-Variable Portfolio Values pre-Covid}
    \label{s1}
\end{figure}

\begin{figure}[h]
    \centering
    \includegraphics[width=0.45\textwidth]{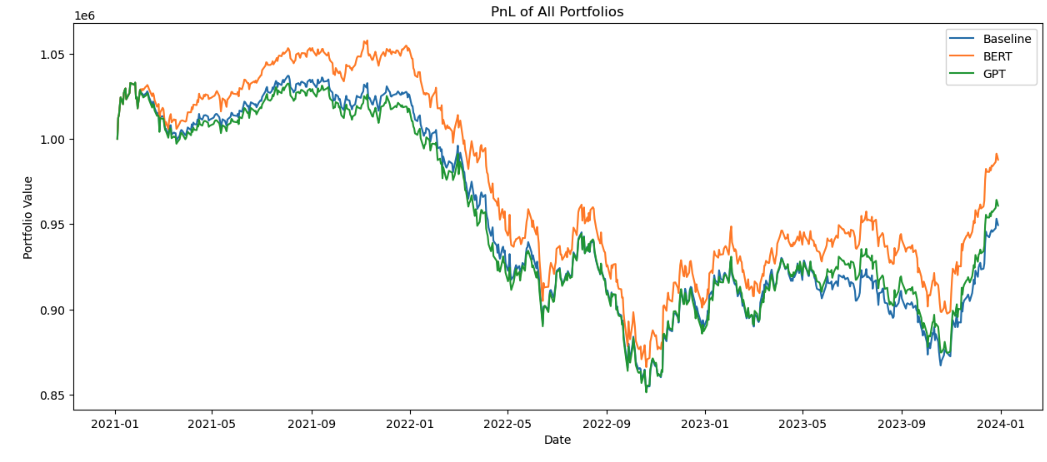}
    \caption{2-Variable Portfolio Values post-Covid}
    \label{s2}
\end{figure}

\begin{table}[h]
\centering
\begin{tabular}{||c c c c||} 
 \hline
 Period & Baseline & BERT & GPT\\ [0.5ex] 
 \hline\hline
 Pre-Covid & 2.605 & 2.768 & 2.334\\ 
 \hline
 Post-Covid & -0.904& -0.549 & -0.802  \\
 \hline
\end{tabular}
\caption{Sharpe Ratios for 2-Variable Portfolios}
\label{t9}
\end{table}

We notice that in both periods, the BERT model tends to perform the best, whereas the GPT model sometimes even performs worse than simple rolling covariances. In this case, it seems that using the BERT model to analyze the Beige Book might yield the most optimal portfolio.

\subsection{Multi-Variable}
\begin{figure}[h]
    \centering
    \includegraphics[width=0.45\textwidth]{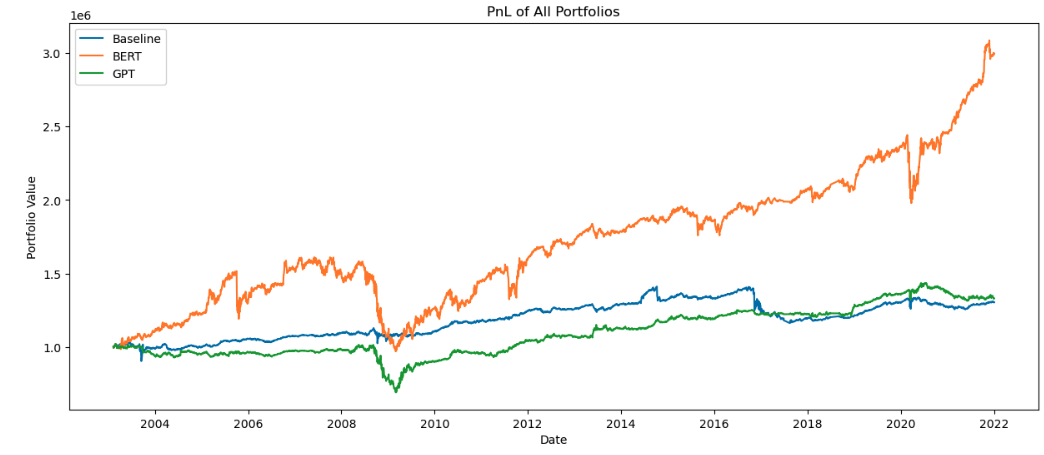}
    \caption{Multi-Variable Portfolio Values pre-Covid}
    \label{m1}
\end{figure}

\begin{figure}[h]
    \centering
    \includegraphics[width=0.45\textwidth]{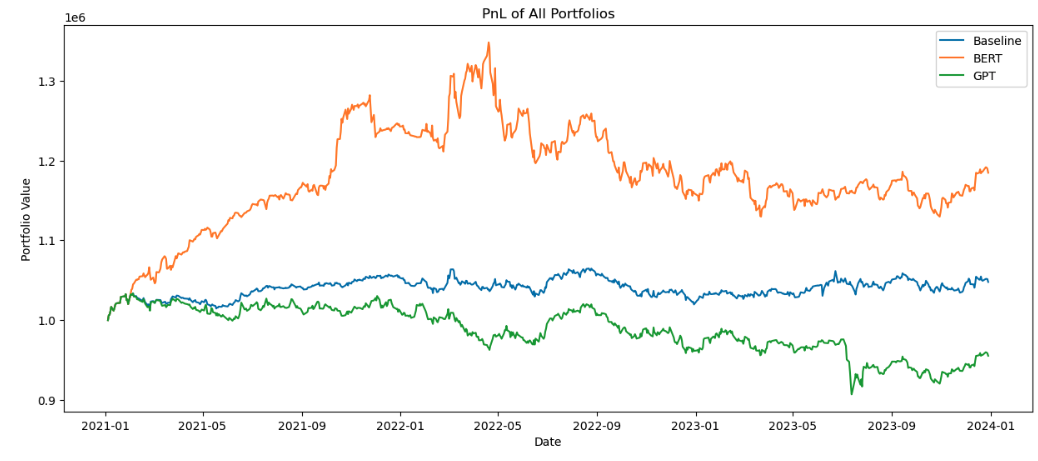}
    \caption{Multi-Variable Portfolio Values post-Covid}
    \label{m2}
\end{figure}

For further testing, we attempt the same simulations except with multiple different variables. We show the portfolio values in figures \ref{m1} and \ref{m2} and the final sharpe ratios in table \ref{t10}.\\

\begin{table}[h]
\centering
\begin{tabular}{||c c c c||} 
 \hline
 Period & Baseline & BERT & GPT\\ [0.5ex] 
 \hline\hline
 Pre-Covid & 0.423 & 1.9208 & 0.445\\ 
 \hline
 Post-Covid & -0.145& 0.790 & -1.009  \\
 \hline
\end{tabular}
\caption{Sharpe Ratios for Multi-Variable Portfolios}
\label{t10}
\end{table}

Here, we notice that the BERT model consistently outperforms both the baseline and the GPT model in both the pre- and post-Covid simulations. Furthermore, it seems that the GPT strategy does not perform significantly better in the pre-covid case, and does even worse after Covid. Hence, this could be further evidence of the GPT model overfitting, or that the GPT does not analyze the Beige Book as well.

\section{Conclusions}
In this report, we show several different ways to compare whether the GPT model provides useful information compared to other traditional models. We demonstrate that the GPT model has evidence of look ahead bias, and that there is no evidence that the GPT model performs that much better in predicting correlations compared to the BERT model. With regard to an optimal strategy, we demonstrate that the BERT model, and sometimes even simply using rolling averages, is a better strategy than asking the GPT model. This is most likely due to the fact that we are treating the BERT model as a classification model, whereas we are still treating the GPT model as a generative model. Hence, classification might be performing better because it being a simpler task. Furthermore, while it is possible that the portfolio performs better with a large number of assets, the number of correlations increases in $O(N^2)$, making it quite expensive to ask GPT for correlations.
\subsection{Further Directions}
There are several directions that could be built on these strategies. For one, while we focused mainly on the Beige Book on the data source, there are other federal data sources that could be utilized, or other cleaned up news data sets that hae been published. Other large language models have also yielded different results, such as the Gemini or Llama models. Even the GPT-4o model could be retested in a few years, given enough data for testing out of sample. Overall, we provide a basis for the different ways of testing how large language models compare to earlier transformer models through predictions on less-utilized variables.

\section*{Acknowledgment}

I would like to thank professor Burton Hollifield for advising me throughout this project and professor Ruben Martins for teaching the course 07-400 and helping throughout Meeting of the Minds.

\clearpage
  \appendices
\onecolumn
  \section{Errors of Experiments Visualized}
  \label{FirstAppendix}
  Below we show the full distribution of errors for the different hypothesis tests described in section \ref{res}. For the sake of clarity, we choose the 3rd version for each model as well as set the number of months to 1, which directly measures the errrors of each month:
\begin{figure}[h]
    \centering
    \begin{minipage}{0.5\textwidth}
        \centering
        \includegraphics[width=\textwidth]{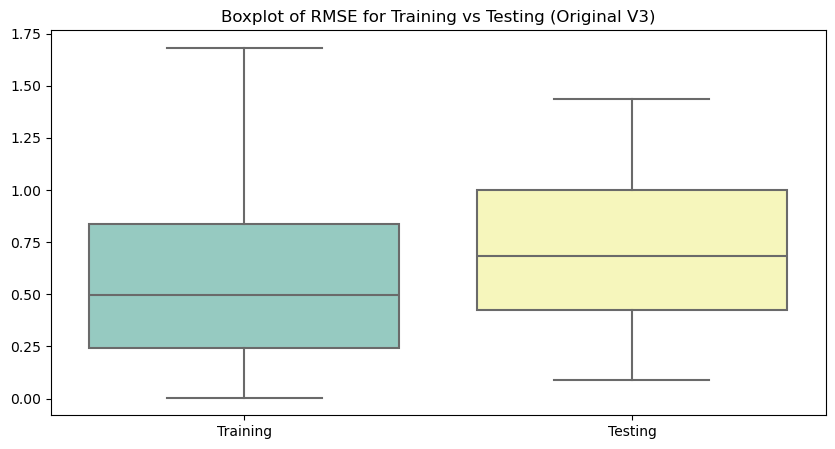}
        \caption{Errors for out-of-sample (Original)}
        \label{s2a}
    \end{minipage}%
    \begin{minipage}{0.5\textwidth}
        \centering
        \includegraphics[width=\textwidth]{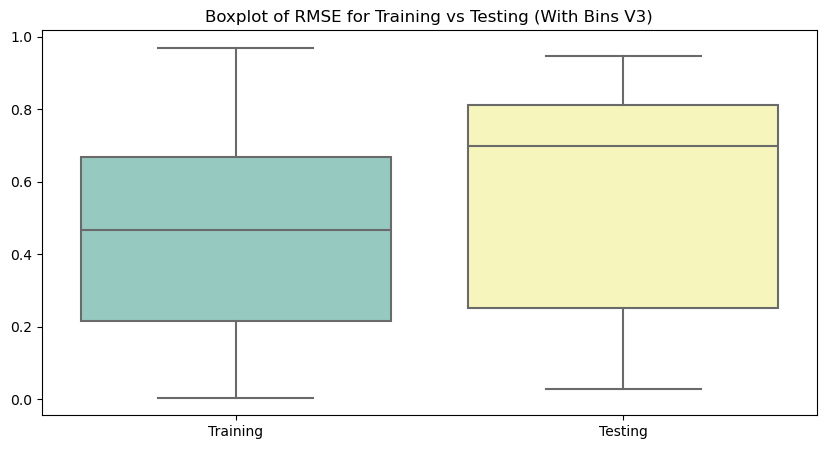}
        \caption{Errors for out-of-sample (With Bins)}
        \label{s2b}
    \end{minipage}
\end{figure}
\begin{figure}[h]
    \centering
    \begin{minipage}{0.5\textwidth}
        \centering
        \includegraphics[width=\textwidth]{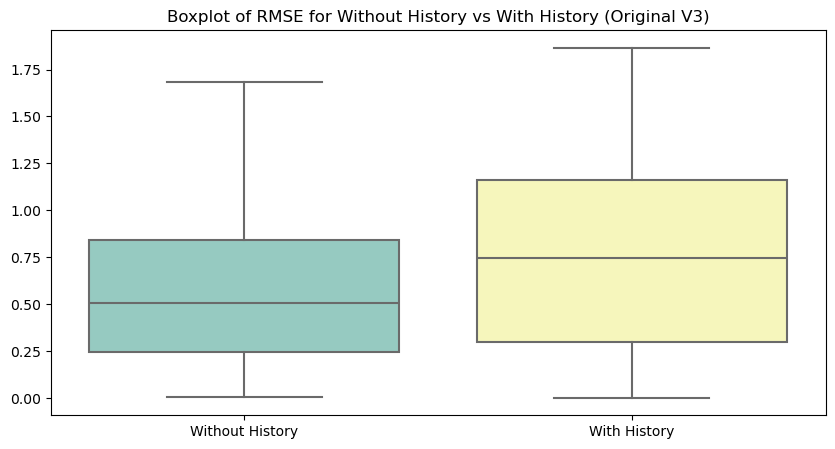}
        \caption{Errors for Adding History (Original)}
        \label{s1}
    \end{minipage}%
    \begin{minipage}{0.5\textwidth}
        \centering
        \includegraphics[width=\textwidth]{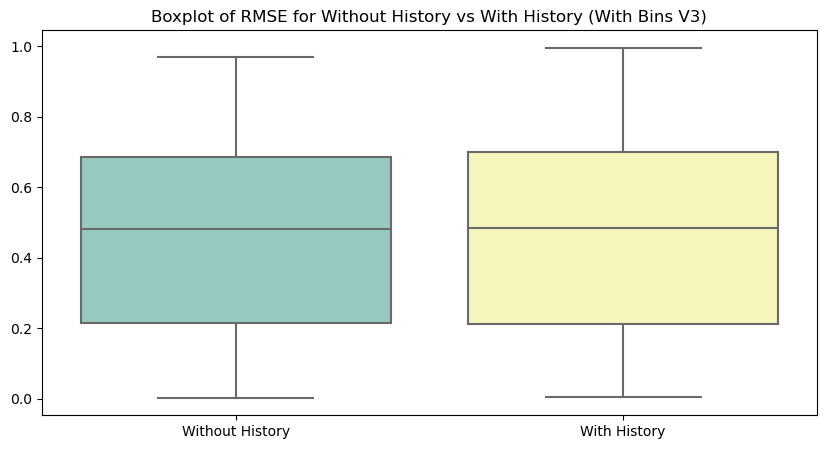}
        \caption{Errors for Adding History (With Bins)}
        \label{s2}
    \end{minipage}
\end{figure}
\begin{figure}[h]
    \centering
    \begin{minipage}{0.5\textwidth}
        \centering
        \includegraphics[width=\textwidth]{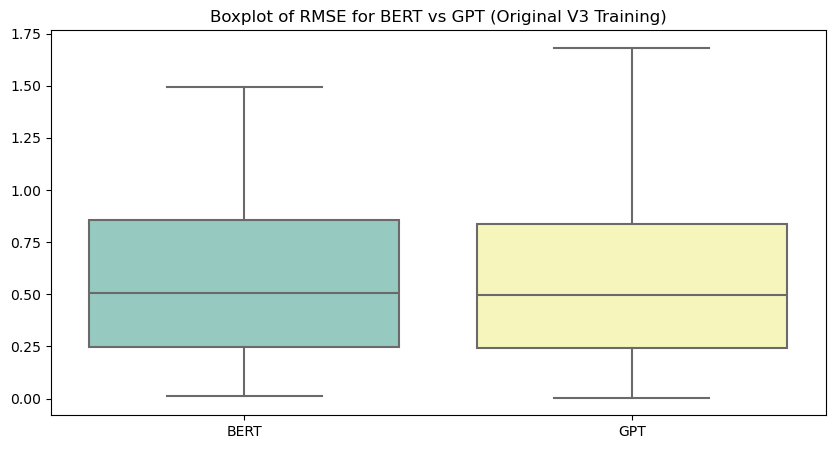}
        \caption{Training Errors for BERT vs GPT (Original)}
        \label{train1}
    \end{minipage}%
    \begin{minipage}{0.5\textwidth}
        \centering
        \includegraphics[width=\textwidth]{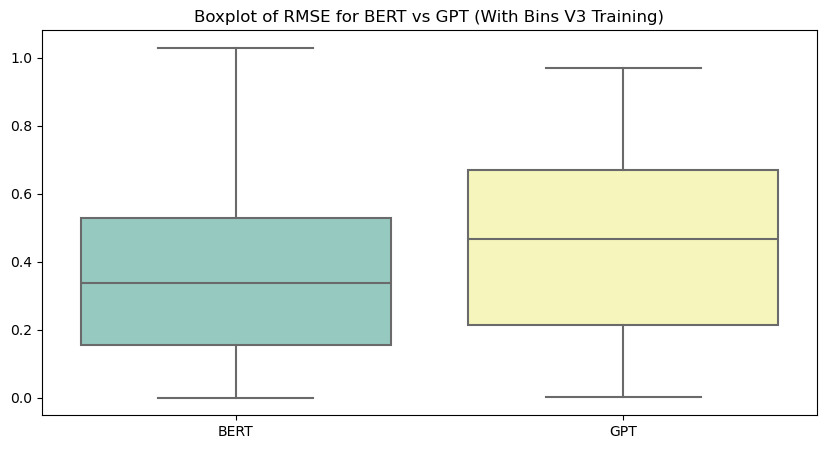}
        \caption{Training Errors for BERT vs GPT (With Bins)}
        \label{train2}
    \end{minipage}
\end{figure}

\begin{figure}[h]
    \centering
    \begin{minipage}{0.5\textwidth}
        \centering
        \includegraphics[width=\textwidth]{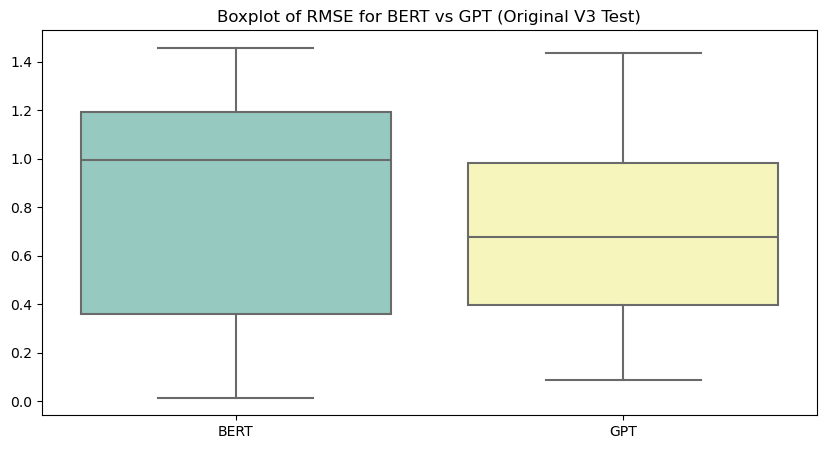}
        \caption{Testing Errors for BERT vs GPT (Original)}
        \label{test1}
    \end{minipage}%
    \begin{minipage}{0.5\textwidth}
        \centering
        \includegraphics[width=\textwidth]{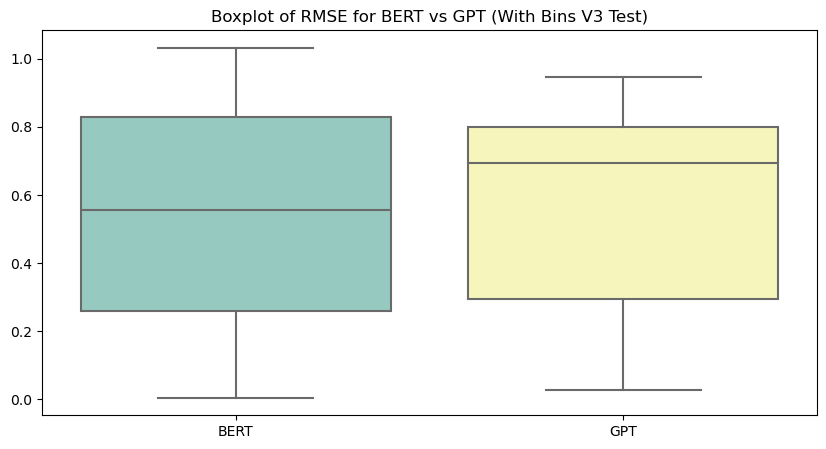}
        \caption{Testing Errors for BERT vs GPT (Original)}
        \label{test2}
    \end{minipage}
\end{figure}

\end{document}